\documentclass[aps,prl, amsmath, showpacs, preprintnumbers,superscriptaddress, twocolumn,sort&compress,floatfix, amssymb]{revtex4}
\pdfoutput=1
\usepackage{graphicx}
\usepackage{rotating}
\usepackage{dcolumn}
\usepackage{bm}
\usepackage{color}
\usepackage{mathptmx, textcomp}
\usepackage[latin1]{inputenc}
\usepackage{braket}

\usepackage{multirow}

\begin{document}

\author{F. Meinert}
\affiliation{Institut f\"ur Experimentalphysik und Zentrum f\"ur Quantenphysik, Universit\"at Innsbruck, 6020 Innsbruck, Austria}
\author{M. J. Mark}
\affiliation{Institut f\"ur Experimentalphysik und Zentrum f\"ur Quantenphysik, Universit\"at Innsbruck, 6020 Innsbruck, Austria}
\affiliation{Institut f\"ur Quantenoptik und Quanteninformation, \"Osterreichische Akademie der Wissenschaften, 6020 Innsbruck, Austria}
\author{K. Lauber}
\affiliation{Institut f\"ur Experimentalphysik und Zentrum f\"ur Quantenphysik, Universit\"at Innsbruck, 6020 Innsbruck, Austria}
\author{A. J. Daley}
\affiliation{Department of Physics and SUPA, University of Strathclyde, Glasgow G4 0NG, UK}
\author{H.-C. N\"agerl}
\affiliation{Institut f\"ur Experimentalphysik und Zentrum f\"ur Quantenphysik, Universit\"at Innsbruck, 6020 Innsbruck, Austria}

\title{Floquet engineering of correlated tunneling in the Bose-Hubbard model with ultracold atoms}

\date{\today}

\pacs{37.10.Jk, 67.85.Hj, 03.75.Lm, 71.45.Gm}

\begin{abstract}
We report on the experimental implementation of tunable occupation-dependent tunneling in a Bose-Hubbard system of ultracold atoms via time-periodic modulation of the on-site interaction energy. The tunneling rate is inferred from a time-resolved measurement of the lattice site occupation after a quantum quench. We demonstrate coherent control of the  tunneling dynamics in the correlated many-body system, including full suppression of tunneling as predicted within the framework of Floquet theory. We find that the tunneling rate explicitly depends on the atom number difference in neighboring lattice sites. Our results may open up ways to realize artificial gauge fields that feature density dependence with ultracold atoms.
\end{abstract}

\maketitle

Ultracold atoms in optical lattices provide an excellent platform for highly controlled studies of strongly correlated states of matter, paradigm examples being the observation of a Mott insulator for bosons \cite{Greiner2002} and fermions \cite{Joerdens08,Schneider08}. At the same time, the unprecedented temporal control over system parameters allows for the investigation of coherent dynamics in the many-body system \cite{Polkovnikov2011,Eisert15}. Among a plethora of associated phenomena, time-periodic driving of the quantum system opens up particularly exciting possibilities. The effect of such a modulation can often be described within an effective time-independent Floquet-Hamiltonian with novel synthetically engineered properties \cite{Goldman14,Bukov15,Holthaus16}. Pioneering work in this direction has demonstrated the coherent control of the single-particle tunneling amplitude in shaken lattices \cite{Lignier07}, with applications towards shifting phase boundaries \cite{Zenesini09,Eckardt05} and transport \cite{Alberti09,Haller10}, observation of magnetic frustration \cite{Struck11}, and even the realization of artificial gauge potentials \cite{Aidelsburger13,Miyake13,Struck13} and topological band structures \cite{Jotzu14}. Here, we go beyond control of the single-particle Hamiltonian by implementing tunable correlated tunneling for which coupling explicitly depends on the presence of other particles in the interacting many-body system \cite{Rapp12}. This opens a new platform for the exploration of phenomena in lattices with occupation-dependent hopping, ranging from a variety of extended Hubbard models with rich phase diagrams for bosons \cite{Dutta15} and fermions \cite{Liberto14,Arrachea94,Arrachea96} to situations with broken mirror symmetry \cite{Greschner14b} and potentially dynamical synthetic gauge fields \cite{Greschner14}.

\begin{figure}
\includegraphics[width=0.48\columnwidth]{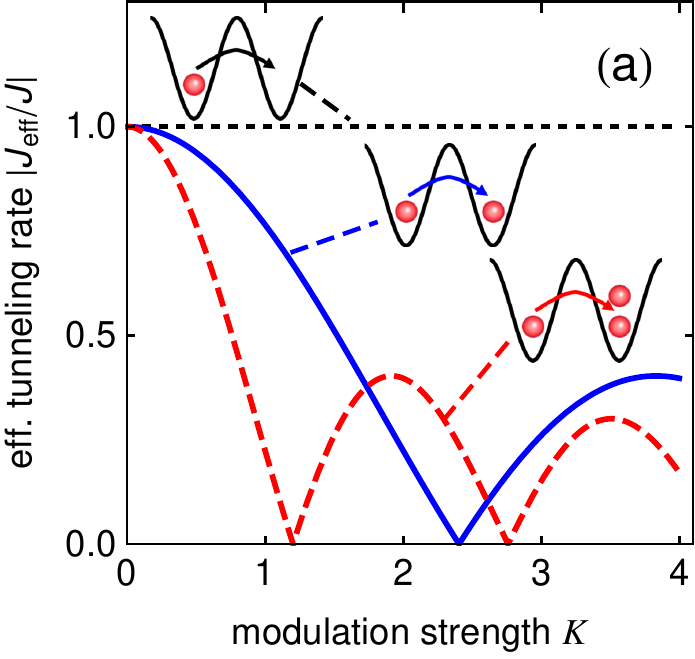}
\hspace{2mm}
\includegraphics[width=0.472\columnwidth]{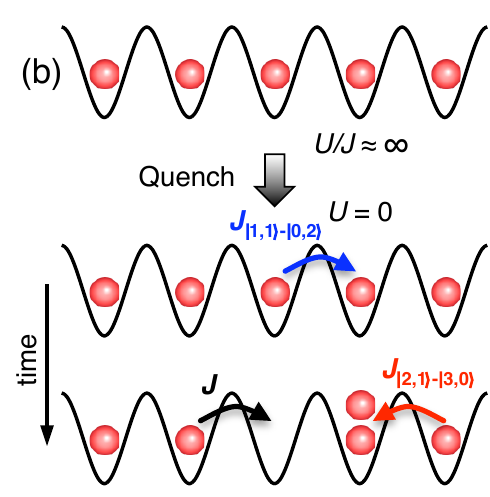}
\caption{\label{FIG1}(color online) Controlling occupation-dependent tunneling with periodically modulated interactions. (a) Renormalized tunneling rate $J_{\rm{eff}}/J$ as a function of the modulation strength $K$ for the processes $|1,1\rangle - |0,2\rangle$ (solid line) and $|1,2\rangle - |0,3\rangle$ (dashed line). Single-particle tunneling ($|1,0\rangle - |0,1\rangle$) is not affected by the modulation (dotted line). (b) Illustration of the experimental measurement protocol. Starting from a Mott insulator with unity filling (top row) the interaction strength is suddenly quenched to zero. The subsequent delocalization of the initially localized atoms, resulting in multiply occupied lattice sites, is controlled via the modulation strength.}
\end{figure}

Our approach exploits a strongly correlated lattice gas of bosonic cesium (Cs) atoms in the presence of periodically modulated particle-interactions, described by the Bose-Hubbard Hamiltonian \cite{Jaksch1998}
\begin{equation}
\hat{H} = -J \sum\limits_{\langle i,j \rangle } \hat{a}_i^\dagger \hat{a}_j + \sum\limits_{i} \frac{U(t)}{2} \hat{n}_i\left(\hat{n}_{i}-1\right) \, .
\label{EQ1}
\end{equation}
As usual, $\hat{a}_i^\dagger$ ($\hat{a}_i$) are the bosonic creation (annihilation) operators at the $i$th lattice site, $\hat{n}_i = \hat{a}_i^\dagger \hat{a}_i$ are the number operators, and $J$ is the tunnel matrix element between neighboring lattice sites. At the heart of the experiments reported in this Letter is a time-dependent, rapidly oscillating on-site interaction energy $U(t)$ of the form $U(t)=U + \delta U \sin \left( 2 \pi f_{\rm{mod}} t \right)$. For a sufficiently large oscillation frequency $h f_{\rm{mod}} \gg U,J$, one obtains an effective time-independent description by expanding (\ref{EQ1}) in a Floquet basis and retaining a single time-averaged Floquet sector \cite{Rapp12}. In striking contrast to shaken lattices, modulated interactions generate dynamics that explicitly depend on the occupation number \cite{Rapp12,Greschner14}, so that the modulation-induced tunneling in the effective Hamiltonian,
\begin{equation}
\hat{H}_{\rm{eff}} = -J \sum\limits_{\langle i,j \rangle } \hat{a}_i^\dagger J_{0}\left( K (\hat{n}_i - \hat{n}_j) \right) \hat{a}_j + \sum\limits_{i} \frac{U}{2} \hat{n}_i\left(\hat{n}_{i}-1\right) \, ,
\label{EQ2}
\end{equation}
is now occupation dependent. Specifically, the amplitudes $J_{\rm{eff}} = J \times J_0(K \Delta n)$ of the tunneling processes are determined by the strength of the modulation $K=\delta U/(h f_{\rm{mod}})$ via a rescaling with the zeroth-order Bessel function $J_{0}$ (Fig.~\ref{FIG1}(a)). One consequence is a suppression of hopping for specific choices of $K$, which explicitly depends on the particle number difference $\Delta n$ (after applying $\hat{a}$) at the lattice sites involved, known as many-body coherent destruction of tunneling \cite{Gong09}.

We demonstrate the modulation-induced coherent control of occupation-dependent atom tunneling after a quantum quench in the correlated many-body system. Starting from localized bosons prepared in an atomic Mott insulator, we suddenly switch off particle interactions by means of a Feshbach resonance and identify the subsequent delocalization processes as a sensitive probe for hopping in the lattice. This forms the basis for a controlled study of the many-body tunneling dynamics in the presence of periodically modulated interactions (Fig.~\ref{FIG1}(b)).

Our experiments start with a Cs Bose-Einstein condensate \cite{Kraemer2004}, from which we prepare a three-dimensional (3D) one-atom-per-site Mott insulator in a cubic optical lattice at a scattering length $a_{\rm{s}} \approx 230 \, a_0$ \cite{supmat}. The lattice depth in all directions is $V_{q} = 20 E_{\rm{R}}$ ($q=x,y,z$), where $E_{\rm{R}}=h \times 1.325 $ kHz is the photon recoil energy. A Feshbach resonance allows us to control $a_{\rm{s}}$ and thereby $U$ independent of $J$ \cite{Mark2012}. First, we investigate tunneling dynamics in the lattice after a sudden quench of $U$ to the vicinity of the non-interacting limit. Specifically, starting deep in the Mott-insulating phase with localized atoms, we quickly ramp $U$ to a value in the range $|U/h| \lesssim 200~\rm{Hz}$ and wait for a hold time of $t_{\rm{h}} = 50~\rm{ms}$. Subsequently, the scattering length is rapidly set back to its initial value, which freezes the local site occupancies. We then detect the total number of atoms residing in singly and doubly occupied lattice sites via Feshbach-molecule formation and detection, involving a narrow g-wave Feshbach resonance with a pole at $19.8$ G \cite{Meinert2013}. The number of atoms in sites occupied by more than two bosons is inferred by means of controlled recombination loss \cite{Campbell06,Jack03}. For this, we quickly ramp the magnetic field to $9.3$ G, where $a_{\rm{s}}$ is large and negative, inducing fast loss due to three-body recombination \cite{Kraemer06}. The sample is held for 10 ms, long enough to lose all particles at sites with occupation number larger than two, before detecting the remaining atoms. Note that the site occupancies are measured individually in independent realizations of the experiment.

\begin{figure}
\includegraphics[width=0.32\columnwidth]{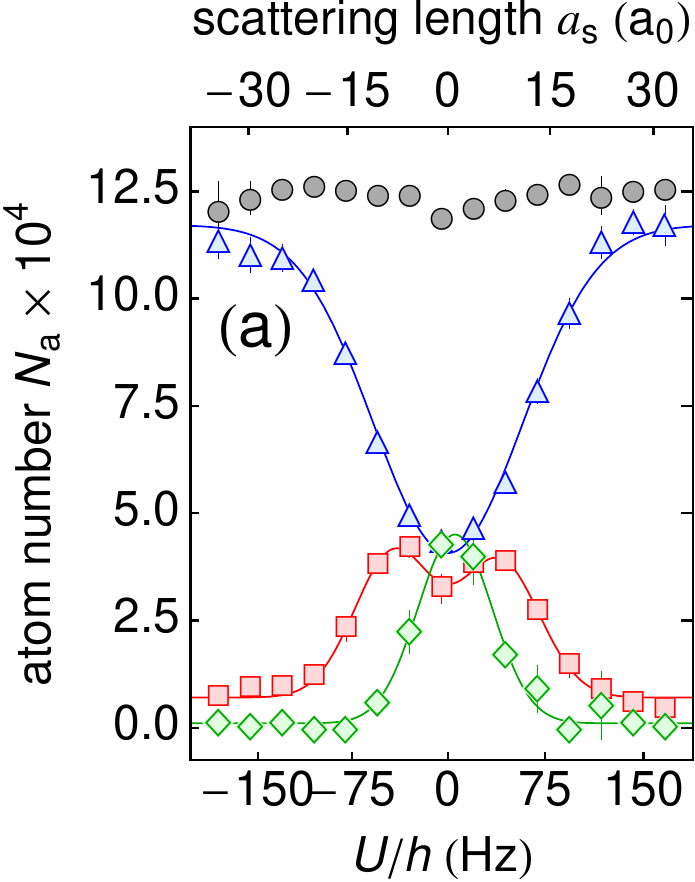}
\includegraphics[width=0.32\columnwidth]{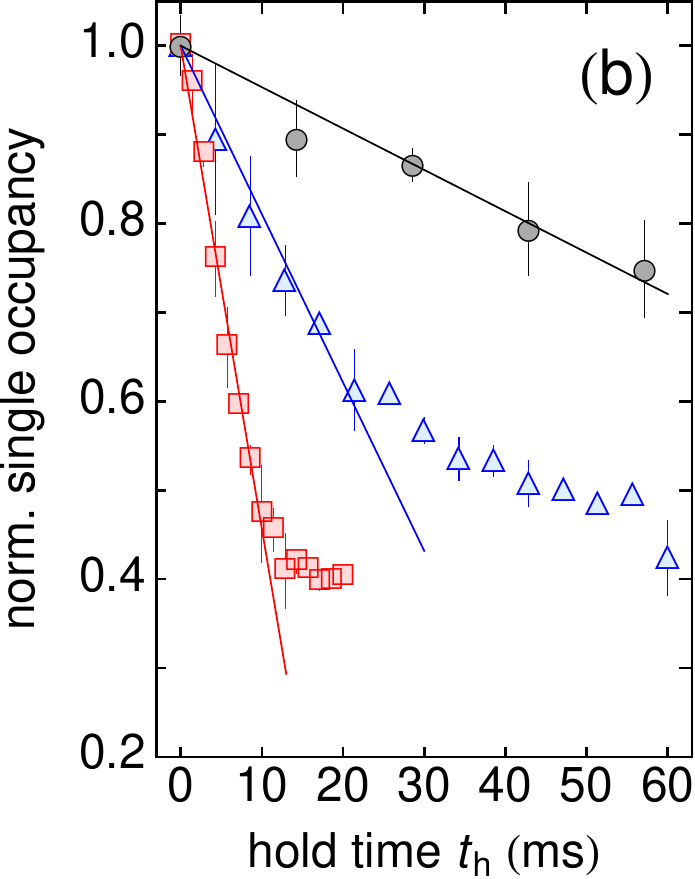}
\includegraphics[width=0.32\columnwidth]{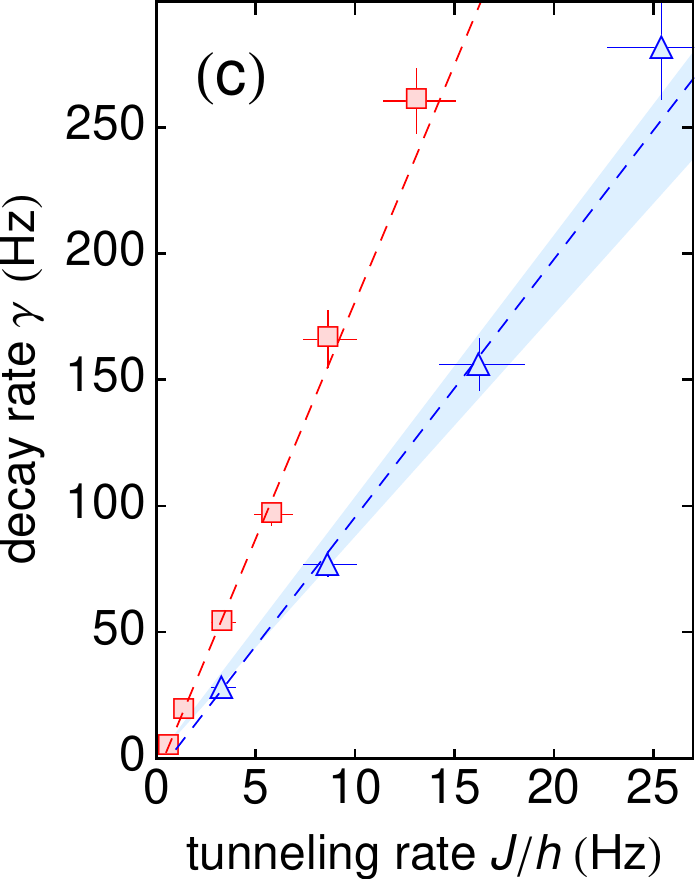}
\caption{\label{FIG2}(color online) Decay of singly occupied lattice sites after a quench from a Mott insulator with unity filling to non-interacting bosons. (a) Number of atoms in singly (triangles) and doubly (squares) occupied sites, in sites with occupation number $\geq 3$ (diamonds), as well as the total number of atoms (circles) as a function of $U$ after a hold time $t_{\rm{h}}=50$ ms for $V_{x,y,z} = 20 E_{\rm{R}}$. Solid lines show (double-)Gaussian fits to the data. (b) Normalized number of singly occupied sites as a function of $t_{\rm{h}}$ after a quench to $U=0$ for $V_{x,y,z} = 20 E_{\rm{R}}$ (squares), $25 E_{\rm{R}}$ (triangles), and $30 E_{\rm{R}}$ (circles). Solid lines show linear fits to the initial decay. (c) Initial decay rate $\gamma$ of singly occupied sites after a quench to $U=0$ as a function of the tunneling rate $J$ in 3D (squares) and 1D (triangles). The dashed lines are linear fits to the data. The shaded area indicates an estimate for $\gamma$ as extracted from numerical simulations in 1D \cite{supmat}.}
\end{figure}

The result of this measurement is shown in Fig.~\ref{FIG2}(a). While the total atom number of the sample (circles) remains constant as a function of $U$, a pronounced resonance for the number of singly occupied lattice sites (triangles) is observed. The position of the resonance minimum is identified with a vanishing value of $U$. A Gaussian fit to the data reveals a half-width-at-half-maximum (HWHM) of 68(3)~Hz, which is comparable with the calculated lattice bandwidth, $12J/h = 39.6~\rm{Hz}$. The reduction of the number of lattice sites with unity filling arises from tunneling-induced delocalization, which is energetically allowed for sufficiently small $U$. That process is accompanied by the build-up of multiply occupied lattice sites. Accordingly, after the time evolution we observe atoms residing in sites occupied by two (squares) and more than two (diamonds) bosons resonantly enhanced around $U=0$. For the latter, note the reduced Gaussian HWHM of $34(2)$ Hz, reflecting the larger on-site energy associated with three particles on the same lattice site.

Next, we study the dynamics after the quench to $U=0$ for different depths of the optical lattice. In order to provide similar starting conditions for all measurements, the initial Mott-insulating state is now prepared in a deeper potential with $V_{x,y,z} = 30 E_{\rm{R}}$. The quench in $U$ is done as before, but now it is accompanied by an additional rapid change of the lattice depth to the desired value. After letting the sample evolve for a variable time $t_{\rm{h}}$, the optical lattice and $a_{\rm{s}}$ are both ramped back to their initial values, followed by the aforementioned detection scheme for determining site occupancies. In Fig.~\ref{FIG2}(b), the normalized number of singly occupied lattice sites as a function of $t_{\rm{h}}$ is depicted for three different values of the lattice depth. An initial decay is observed over a timescale that is on the order of the single particle tunneling time, before the signal levels off to a stationary value. The initial decay is quantified by a linear fit, the slope of which delivers a characteristic decay rate $\gamma$. In Fig.~\ref{FIG2}(c), we show values for $\gamma$ obtained from such measurements at different lattice depths as a function of the corresponding calculated tunneling rate $J$ (squares). The data suggest a linear dependence, which invites us to exploit the decay of singly occupied sites after the quench as a measure for the tunneling rate in the lattice.

Similarly, we measure the decay of single occupancy in 1D chains by varying only the vertical lattice depth $V_{z}$ while keeping $V_{x,y}$ fixed at $30 E_{\rm{R}}$. Here, $V_z$ is kept small compared to $V_{x,y}$ to ensure decoupling of the 1D systems over relevant experimental timescales. The results obtained for $\gamma$ as a function of $J$ (triangles in Fig.~\ref{FIG2}(c)) again obey a linear dependence. A quantitative estimate for $\gamma$ as extracted from numerical simulations of a single 1D Bose-Hubbard chain \cite{supmat} is found to be in good agreement with the data. We note that for a fixed $J$ we observe a faster decay of single occupancy in 3D compared to the 1D case, owing to the increased number of nearest neighbors. This is quantified by the ratio of the slopes $\beta_{\rm{3D}}/\beta_{\rm{1D}} = 1.8(1)$, which we obtain from linear fits to the data. The observed ratio is different from $3$, as one may expect from the delocalization process of a single atom by counting nearest neighbors. This we attribute to the presence of many particles that all take part simultaneously in the tunneling dynamics. Indeed, a numerical study of a simplified model with multiple atoms in a 3D and 1D configuration predicts a ratio of $1.6$ in reasonably good agreement with the measured value \cite{supmat}.

\begin{figure}
\includegraphics[width=1\columnwidth]{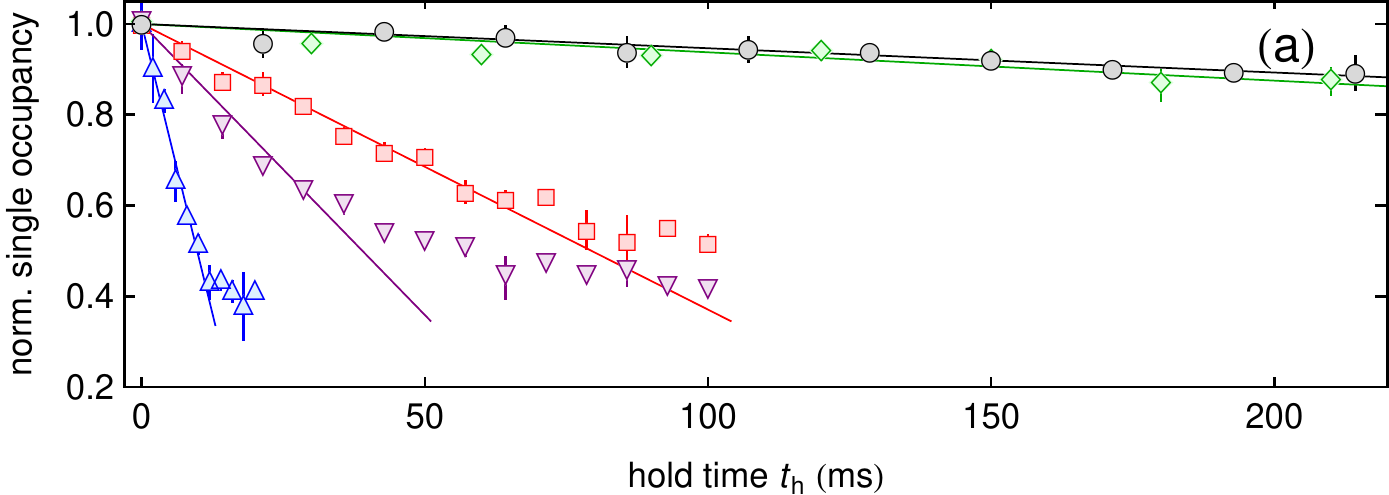}\\
\vspace{2mm}
\includegraphics[width=1\columnwidth]{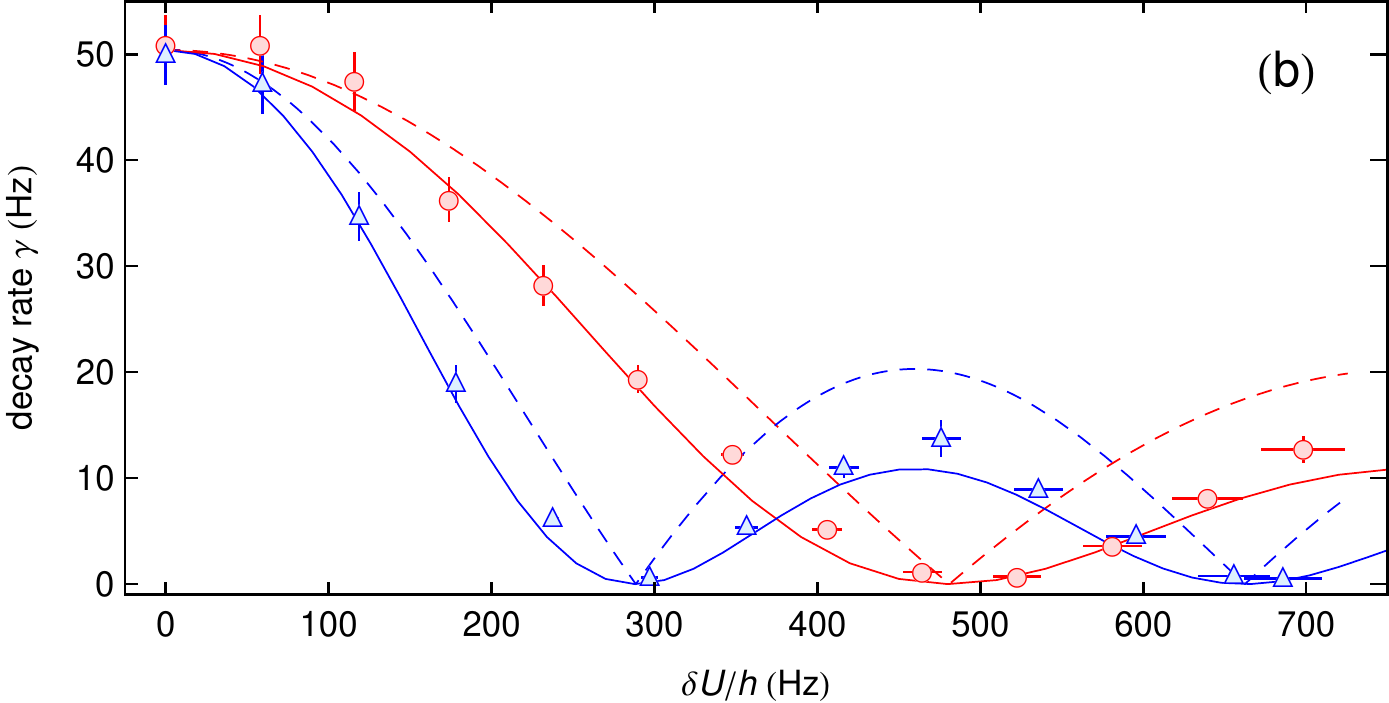}
\caption{\label{FIG3}(color online) Controlling the decay of single occupancy in the presence of modulated interactions after quenching from a Mott insulator with unity filling to $U=0$. (a) Normalized number of singly occupied lattice sites as a function of $t_{\rm{h}}$ after the quench for modulation amplitudes $\delta U/h = 0$ Hz (triangles), 238(3) Hz (squares), 297(4) Hz (diamonds), 476(11) Hz (inverted triangles), and 686(23) Hz (circles) when modulating with $f_{\rm{mod}}=120$ Hz. Solid lines show linear fits to the initial decay. (b) Initial decay rate $\gamma$ of singly occupied sites as a function of $\delta U$ for $f_{\rm{mod}}=120$ Hz (triangles) and 200 Hz (circles). The dashed lines depict $\gamma_{\delta U=0} \times |J_0(\delta U/(h f_{\rm{mod}}))|$. The solid lines show numerical results for $\gamma$ as extracted from a calculation within a 3D lattice of 7 sites in star-type configuration \cite{supmat}. For all measurements in this figure $V_{x,y,z} = 20 E_{\rm{R}}$.}
\end{figure}

Having discussed the bare tunneling dynamics in the many-body system for vanishing particle interaction, we are now in a position to investigate the situation in the presence of a periodically modulated on-site interaction. Starting again from the Mott insulator at $V_{x,y,z} = 20 E_{\rm{R}}$, we perform the quench to $U=0$ as before but now apply an additional sinusoidal modulation with amplitude $\delta U$ and frequency  $f_{\rm{mod}}=120$ Hz during $t_{\rm{h}}$ by a periodic oscillation of the magnetic offset field that controls $a_{\rm{s}}$. The normalized number of singly occupied lattice sites as a function of $t_{\rm{h}}$ is shown in Fig.~\ref{FIG3}(a) for five data sets recorded at increasing values of $\delta U$. For zero modulation strength (triangles) we recognize the decay of sites with unity filling as observed before. Increasing $\delta U$ leads to a substantially slower decay (squares), indicative of a reduced tunneling-induced delocalization of the particles. When further increasing the modulation strength to $\delta U/h = 297(4)$ Hz, the decay is heavily suppressed (diamonds), and essentially allows for a controlled pinning of the particles on their individual lattice sites. For even larger modulation strength, we observe again decay on a comparatively rapid timescale (inverted triangles), while a second strong suppression of tunneling is seen for $\delta U/h = 686(23)$ Hz (circles). We stress that  the lattice depth is kept at a fixed value for all measurements, and tunneling is thus fully controlled via the modulation of particle interactions.

As before, the data are analyzed via a linear fit to the initial decay. From such measurements, we obtain values for the characteristic decay rate $\gamma$ over a wide range of modulation strengths, and plot them in Fig.~\ref{FIG3}(b) as a function of $\delta U$ for two different values of $f_{\rm{mod}}$. We compare the data to the Bessel-function scaling of $|J_{\rm{eff}}|$ within the time-independent Floquet-Hamiltonian description of the system (dashed lines). Clear minima in the measured values for $\gamma$ are found in accordance with the zeros of $J_0(\delta U/(h f_{\rm{mod}}))$. This unequivocally demonstrates a controlled coherent destruction of tunneling in the interacting many-body system that entirely relies on the presence of particles in adjacent lattice sites.

Away from these minima, the experimental data lie significantly below a pure rescaling of $\gamma$ with the Bessel function $J_{0}$. Given the dynamical generation of site occupancies larger than two after the quench in combination with the occupation dependence of $J_{\rm{eff}}$, one may indeed expect a more sophisticated behavior for the decay of single occupancy in the presence of modulated interactions. We attempt to model these effects via a numerical simulation of the time evolution under the action of $\hat{H}_{\rm{eff}}$ within a reduced 3D lattice that consists of 7 sites arranged in a star-type configuration \cite{supmat}. The analysis of single occupancy at the central lattice site delivers a measure for the rate of decay, which reproduces the scaling behavior of the measured $\gamma$ as a function of $\delta U$ (solid lines in Fig.~\ref{FIG3}(b)).

\begin{figure}
\includegraphics[width=1\columnwidth]{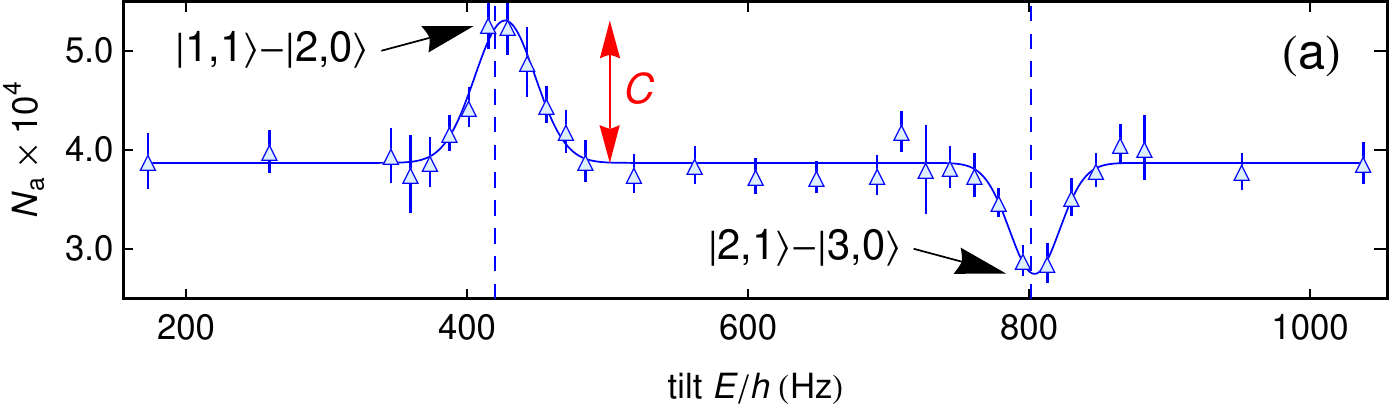}\\
\vspace{2mm}
\includegraphics[width=1\columnwidth]{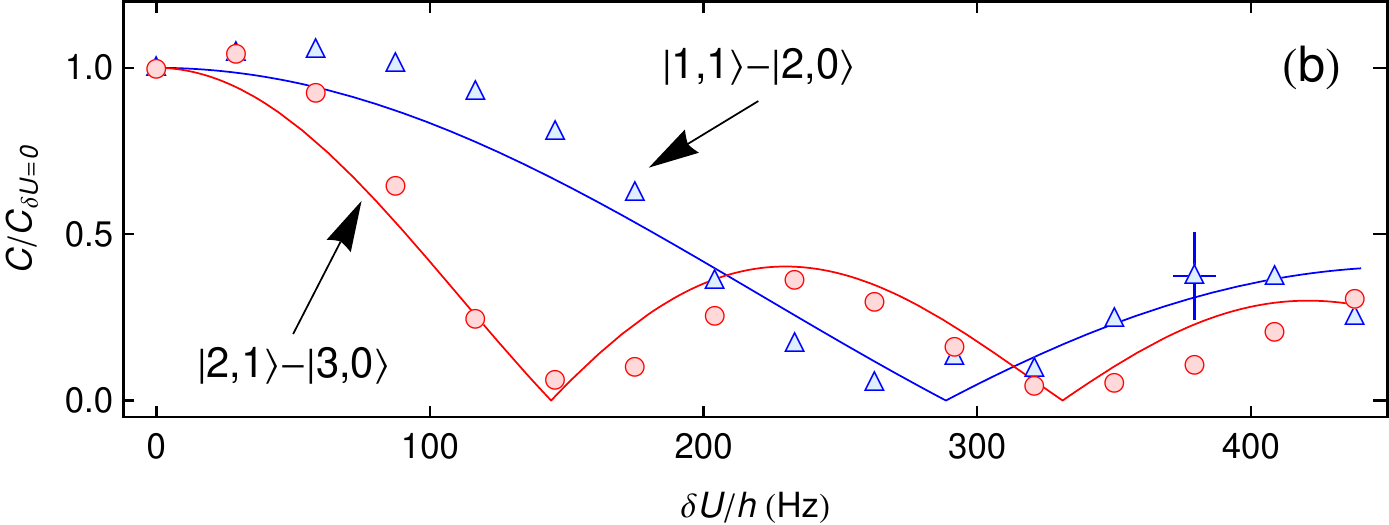}
\caption{\label{FIG4}(color online) Modulation control of occupation-dependent tunneling in the tilted lattice. (a) Number of atoms in doubly occupied lattice sites as a function of the tilt $E$ for $a_{\rm{s}} = 80 \, a_0$ and $V_{x,y,z} = 20 E_{\rm{R}}$. The solid line is a double-Gaussian fit to the data. Vertical dashed lines mark the calculated values for $U$ and $2U$, the latter corrected for multibody interactions \cite{Mark2012,Buechler10}. (b) Normalized resonance peak values $C / C_{\delta U=0}$ for the processes $|1,1\rangle - |2,0\rangle$ (triangles) and $|2,1\rangle - |3,0\rangle$ (circles) as a function of $\delta U$ for $f_{\rm{mod}}=120$ Hz. The solid lines show $|J_0(\Delta n \times \delta U/(h f_{\rm{mod}}))|$ with $\Delta n=1$ and $\Delta n=2$, respectively. Typical error bars are given for the datapoint at $380$ Hz.}
\end{figure}

The occupation dependence of $J_{\rm{eff}}$ is directly observable in a further experiment for that we initially prepare a sample of randomly distributed singly and doubly occupied sites at $U/h=420(20)$ Hz \cite{supmat}. We rapidly apply a linear energy offset $E$ per site along the vertical $z$-direction by a magnetic force, thereby tilting the lattice. After holding the sample for $t_{\rm{h}} = 50$ ms in the tilted configuration, we set $E$ back to zero and detect double occupancy. The number of atoms in doubly occupied sites as a function of $E$ (Fig.~\ref{FIG4}(a)) exhibits two resonances corresponding to resonant tunneling processes of the type $|1,1\rangle - |2,0\rangle$ ($E \approx U$) and $|2,1\rangle - |3,0\rangle$ ($E \approx 2U$) \cite{Meinert2013,Meinert2014}. We now adjust $E$ to either of the detected resonance positions, thereby selectively addressing one of the two tunneling processes, and additionally modulate $U$ during $[0,t_{\rm{h}}]$. The increase (decrease) of double occupancy $C$ normalized to its value without modulation is shown in Fig.~\ref{FIG4}(b) as a function of $\delta U$. The data exhibit clear minima in accordance with the zeros of $J_{\rm{eff}}$, indicating many-body coherent destruction of tunneling. Moreover, the measurement results follow the qualitative trend expected from the Bessel-function scaling of $J_{\rm{eff}}$.

\begin{figure}
\includegraphics[width=1\columnwidth]{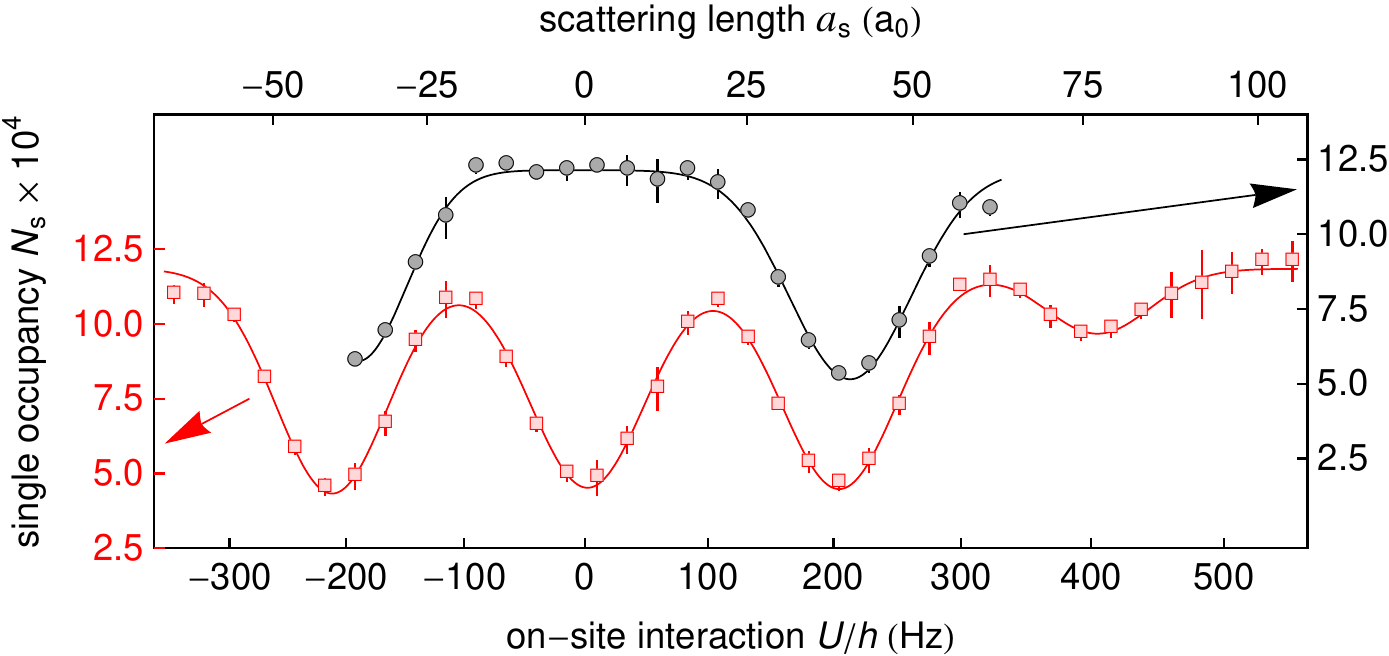}
\caption{\label{FIG5}(color online) Off-resonant and on-resonant tunneling control in the presence of modulated interactions. Number of singly occupied sites as a function of $U$ after $t_{\rm{h}}=50$ ms of modulation at frequency $f_{\rm{mod}}=200$ Hz with amplitudes $ \delta U/h = 290(4)$ Hz (squares) and 523(14) Hz (circles). Here, the lattice depth $V_{x,y,z} = 20 E_{\rm{R}}$. Solid lines show fits to the data using a sum of multiple Gaussians. The data sets are vertically offset for clarity.}
\end{figure}

Finally, we investigate the system at unity filling with modulated interactions at non-zero values of $U$. Again, we start the experiment from initially localized atoms prepared in a Mott insulator with $V_{x,y,z} = 20 E_{\rm{R}}$. However, the Mott insulator is now quenched to a finite $U$ in the range $-400~{\rm{Hz}} \lesssim U/h \lesssim 600~{\rm{Hz}}$. As above, we wait for $t_{\rm{h}}=50$ ms during which the interaction strength is modulated around $U$, before we detect the number of remaining singly occupied lattice sites. The detected atoms residing in unity filled sites as a function of $U$ is shown in Fig.~\ref{FIG5} for two different modulation strengths $\delta U$. Depending on the value of $\delta U$, the range of data shown is restricted by the presence of additional narrow Feshbach resonances that are otherwise crossed during a modulation cycle \cite{supmat}.

We observe a pronounced resonance centered around zero on-site interaction energy when $\delta U/h = 290(4)$ Hz, indicative of the tunneling-induced decay of single occupancy. The absence of this resonance for $\delta U/h = 523(14)$ Hz, which is in the vicinity of the Bessel function zero crossing (cf. Fig.~\ref{FIG3}), reveals the persistence of the coherent destruction of tunneling over an extended range of non-zero interaction strengths. With increasing $U$, the requirement of rapid modulation $h f_{\rm{mod}} \gg U,J$ for the Floquet analysis leading to $\hat{H}_{\rm{eff}}$ breaks down. This is ultimately signaled by resonant formation of particle-hole pairs in the Mott insulator as observed when $U/h\approx f_{\rm{mod}}$ and $U/h\approx 2 f_{\rm{mod}}$. Modulation of $U$ in this resonant regime \cite{Dirks15} thus provides a novel alternative to the more conventional lattice-depth-modulation technique for detecting the Mott gap \cite{Stoeferle04}.

In summary, we have demonstrated tunable occupation-dependent tunneling in a Bose-Hubbard system via periodically driven particle interactions. Our work provides a basis for future studies of interesting many-body phases in correlated hopping models, including insulators with both parity- and string-order \cite{Greschner14b,Endres11,Omran15}. While some of them are expected in the parameter range studied here \cite{Rapp12}, the applicability of our technique can be extended exploiting narrower Feshbach resonances and faster modulation \cite{Clark15}. Moreover, combining modulated interactions with Raman-assisted hopping may allow for the creation of density-dependent synthetic gauge fields \cite{Greschner14,Keilmann11}. This work can also be extended to investigate the phase diagram of recently discussed intermediate-time steady states in driven lattice systems \cite{Bukov15b}.

We are indebted to R. Grimm for generous support. We gratefully acknowledge funding by the European Research Council (ERC) under Project No. 278417 and by the Austrian Science Foundation (FWF) under Project No. I1789-N20. Work at Strathclyde was supported by AFOSR grant FA9550-12-1-0057.

\bibliographystyle{apsrev}

\newpage
\clearpage

\section{Supplementary Material: Floquet engineering of correlated tunneling in the Bose-Hubbard model with ultracold atoms}

\begin{figure}
\includegraphics[width=1\columnwidth]{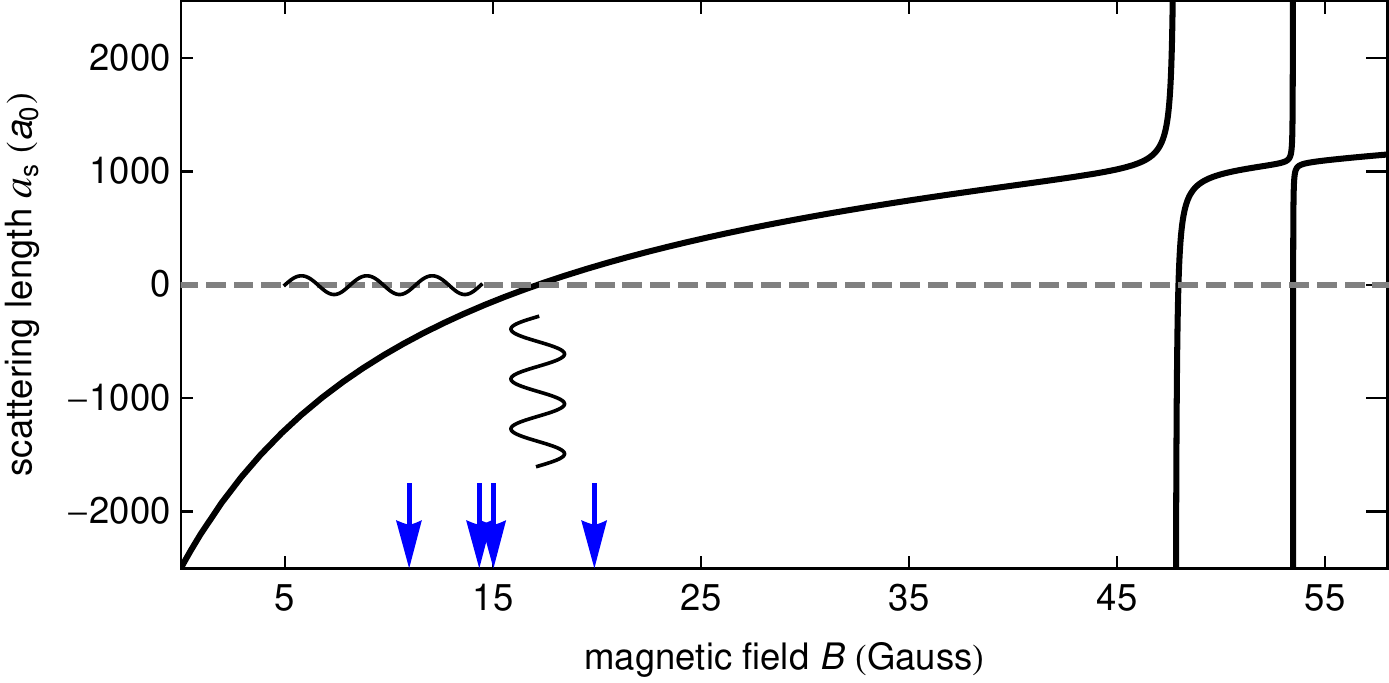}
\caption{\label{FIG6} S-wave scattering length for Cs in its absolute hyperfine ground state $|F=3,m_{F}=3\rangle$ as a function of magnetic field $B$. Modulation of $B$ causes a modulation of $a_{\rm{s}}$ and thereby $U$, as indicated by the sinusoids. Arrows mark the positions of additional narrow Feshbach resonances \cite{Chin2004MAT}.}
\end{figure}

\subsection{Preparation of the one-atom-per-site Mott insulator}

For the preparation of the one-atom-per-site Mott insulator we start with a Bose-Einstein condensate (BEC) of typically $1.1 \times 10^5$ Cs atoms prepared in the internal hyperfine ground state $|F=3,m_{F}=3\rangle$, and held in a crossed beam optical dipole trap. Trapping and cooling procedures are described in Refs.~\cite{Weber2002MAT,Kraemer2004MAT}. The BEC is levitated against gravity by a vertical magnetic field gradient of $\nabla B \approx 31.1$ G/cm. We adiabatically load the sample into a cubic optical lattice generated from three mutually orthogonal retro-refected laser beams at a wavelength $\lambda = 1064.5$ nm, thereby inducing the phase transition to a 3D Mott insulator. The final lattice depth is $V_{q} = 20 E_{\rm{R}}$ in all three directions ($q=x,y,z$), where $E_{\rm{R}}=h^2/(2m\lambda^2) = h \times 1.325$ kHz is the photon-recoil energy with the mass  $m$ of the Cs atom. The scattering length during lattice loading is $a_{\rm{s}} \approx 230 \, a_0$. In the lattice, a broad Feshbach resonance with a pole at $\approx -12$ G allows us to control $a_{\rm{s}}$ and thereby $U$ from large attractive to large repulsive values, including the case of vanishing interactions (see Fig.~\ref{FIG6}) \cite{Mark2011MAT,Mark2012MAT}.

\subsection{Preparation of randomly distributed singly and doubly occupied lattice sites}

For the preparation of a sample of randomly distributed singly and doubly occupied lattice sites, we start from the one-atom-per-site Mott insulator described above. We then quickly ramp $U$ close to the non-interacting limit and wait for 20 ms in order to create double occupancy. Note that $U$ is kept slightly detuned from zero in order to suppress the generation of triply occupied sites. Subsequently, we ramp the interaction strength to $U = 420(20)$ Hz, where we then perform the experiments in the tilted lattice configuration presented in the main article.

\subsection{Modulation of the on-site interaction energy $U$}

We modulate the on-site interaction energy $U$ by a sinusoidal modulation of a magnetic offset field that controls the scattering length $a_{\rm{s}}$ using a broad Feshbach resonance with a pole at $\approx -12$ G. For small amplitudes, the field modulation causes a clean sinusoidal modulation of $a_{\rm{s}}$, as depicted in Fig.~\ref{FIG6}. The amplitude of the modulation is limited by additional narrow Feshbach resonances located at $\approx 15.0$ G and $\approx 19.8$ G \cite{Chin2004MAT}.

\subsection{Numerical simulations within the 1D Bose-Hubbard model}

\begin{figure}
\includegraphics[width=0.48\columnwidth]{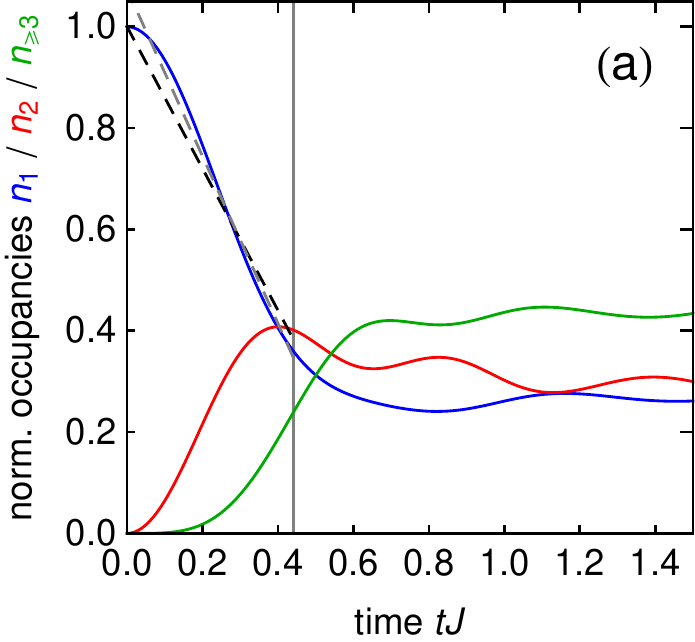}
\hspace{2mm}
\includegraphics[width=0.48\columnwidth]{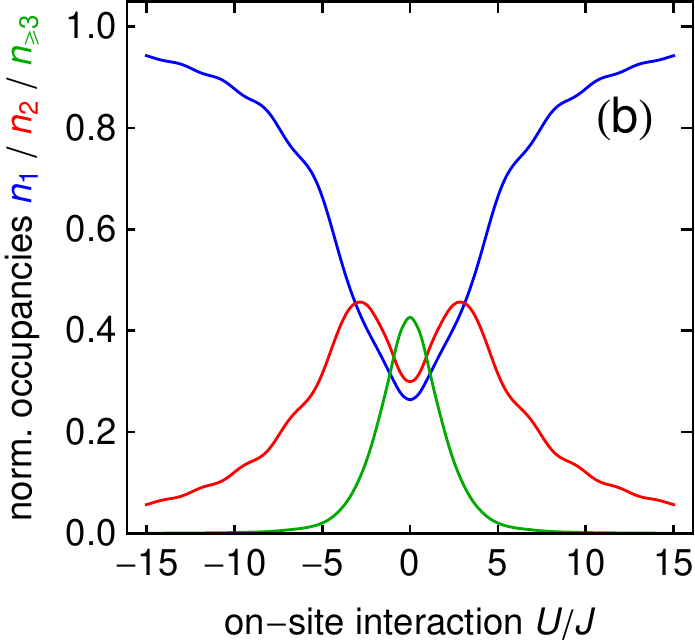}
\caption{\label{FIG7} Numerical simulation for the dynamics of local site occupancies at $U/J$ close to the non-interacting limit in a 1D Bose-Hubbard model starting from an initial one-atom-per-site Fock state. (a) Fraction of atoms $n_{i}$ in lattice sites occupied by $i=1$ (blue), $i=2$ (red), and $i \geq 3$ (green) particles as a function of time for $U/J=0$ ($\hbar=1$). The calculation is done with $N=7$ atoms on $L=7$ lattice sites. Dashed lines show linear fits to the data for $n_1$ to extract an estimate for the initial decay rate $\gamma$. The vertical line denotes the time $t_{\rm{max}}$ up to which the data are fit. (b) Long-time average for $n_1$ (blue), $n_2$ (red), and $n_{\geq 3}$ (green) as a function of $U/J$ ($N=L=7$).}
\end{figure}

For the 1D Bose-Hubbard systems, we compare the measured initial decay rate $\gamma$ of singly occupied lattice sites for $U=0$, presented in Fig.~2(c) of the main article, to an estimate extracted from numerical simulations. For this, we compute the time evolution of local site occupancies within the 1D Bose-Hubbard model by exact diagonalization of the Hamiltonian. Specifically, starting from an initial Fock state with one atom on each site, the time evolution for the fraction of particles $n_i$ in sites occupied by $i$ bosons is derived. The result for $U=0$, depicted in Fig.~\ref{FIG7}(a), shows the decay of $n_1$ together with the build-up of $n_2$. The fraction of particles at sites occupied by three and more bosons are summed up and labeled $n_{\geq 3}$. These calculations are performed for a system of $N=7$ particles on $L=7$ lattice sites.

We extract an estimate for the measured $\gamma$ from linear fits to the initial decay of the numerical data for $n_1$. Fixing the offset of the fit function to unity at $t=0$ or leaving it as a free fit parameter yields reasonable lower and upper bounds for $\gamma$ $\left(1.4 \cdot 2\pi \, J/h \leq \gamma \leq 1.65 \cdot 2\pi \, J/h \right)$. Those define the shaded region shown in Fig.~2(c) of the main article. We have checked for convergence of the extracted values for the decay rate with system size. For completeness, note the difference when comparing the dynamics in the Bose-Hubbard chain to the time evolution for $n_1$ in a double-well system ($N=L=2$). Here, $n_1$ exhibits Rabi oscillations at frequency $f_{\rm{DW}} = 4 J / h$.

Further, we investigate the long-time average of the site occupancies as a function of the interaction strength. For this, we calculate the dynamics at finite $U/J$ analogously to the one shown in Fig.~\ref{FIG7}(a). The time-averaged values for the $n_i$ are computed over the range $0.7 \leq Jt \leq 3$ and plotted in Fig.~\ref{FIG7}(b). The numerical data show qualitatively similar features as the measurements in the 3D system presented in Fig.~2(a) of the main article.

\subsection{Numerical simulations within a simplified 3D Bose-Hubbard model including occupation-dependent hopping}

\begin{figure}
\includegraphics[width=0.48\columnwidth]{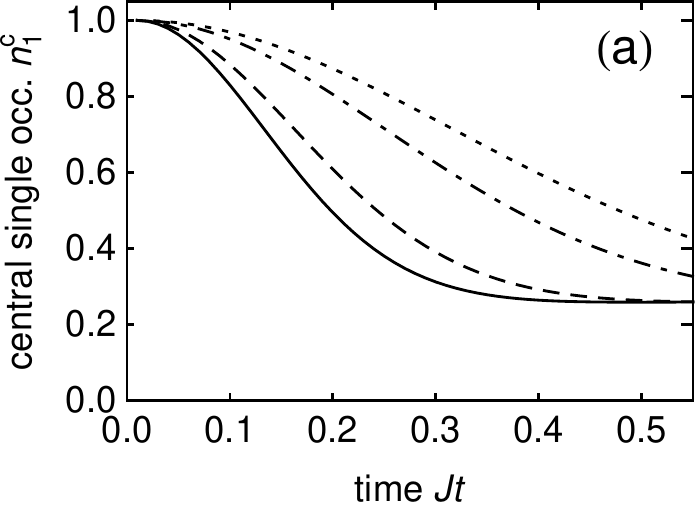}
\hspace{2mm}
\includegraphics[width=0.48\columnwidth]{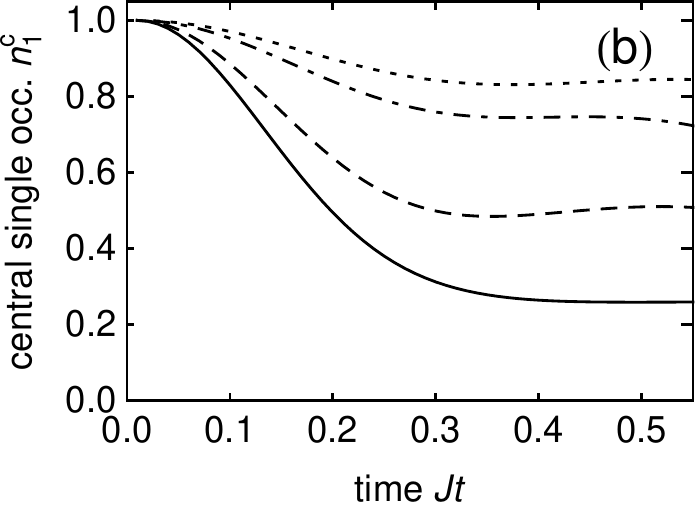}\\
\vspace{2mm}
\includegraphics[width=0.48\columnwidth]{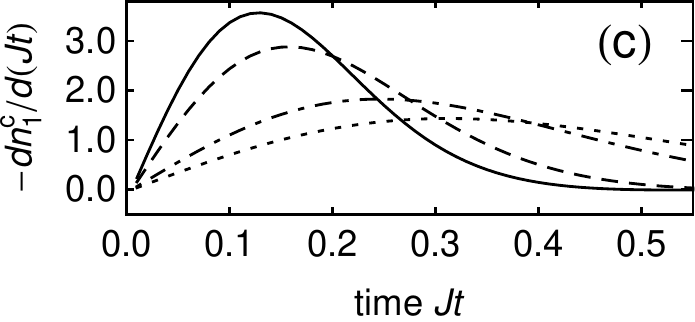}
\hspace{2mm}
\includegraphics[width=0.48\columnwidth]{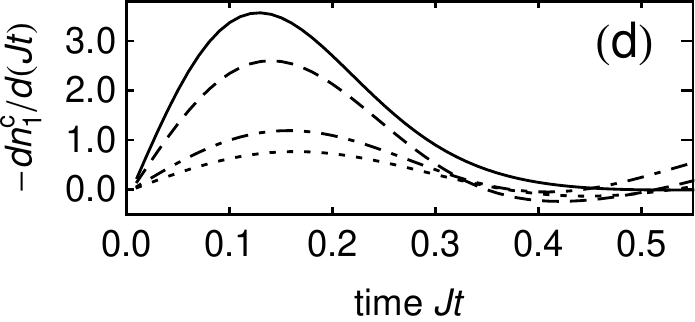}\\
\vspace{2mm}
\includegraphics[width=0.48\columnwidth]{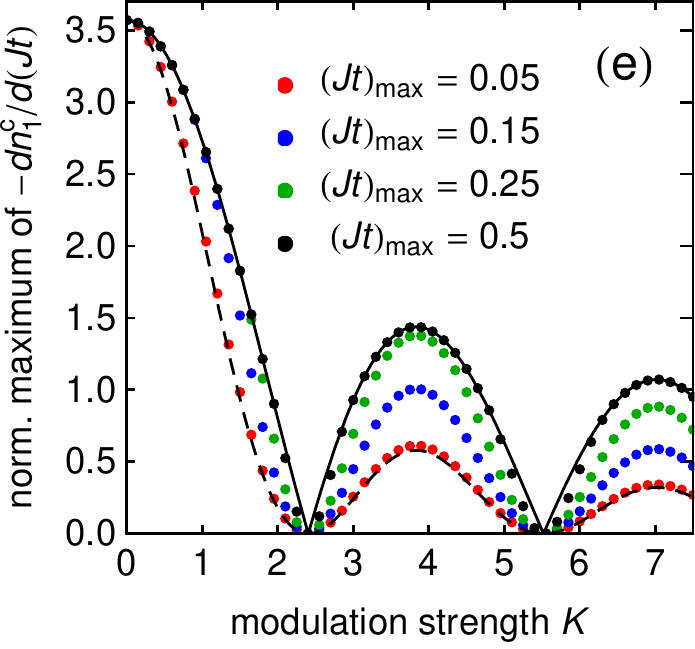}
\hspace{2mm}
\includegraphics[width=0.48\columnwidth]{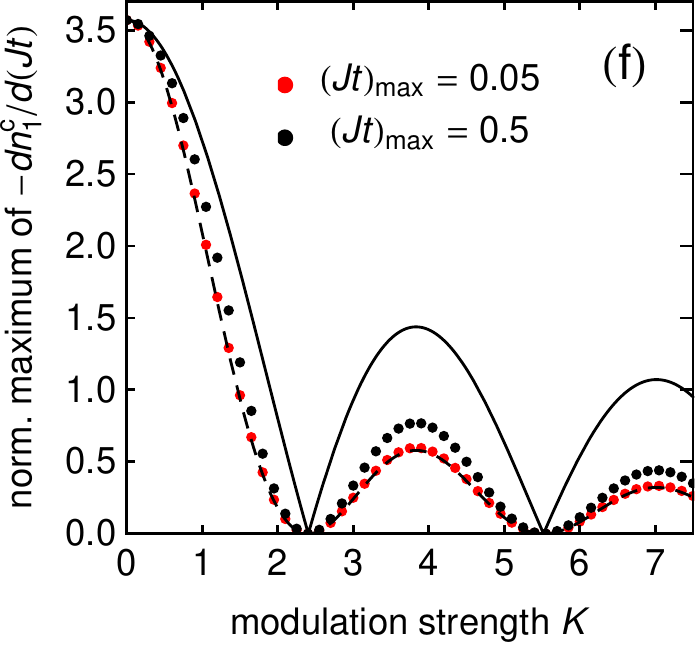}
\caption{\label{FIG8} Numerical simulation for the dynamics of single occupancy at $U=0$ within a 3D star-lattice of $7$ sites. The case of an occupation-independent tunneling rate rescaled $\propto J_{0}(K)$ (left column) is contrasted to the case of the occupation-dependent tunneling in $\hat{H}^{\star}_{\rm{eff}}$ (right column). (a,b) Probability for single occupancy $n_{1}^{\rm{c}}$ in the central site of the \textit{star}-lattice as a function of time starting from an initial one-atom-per-site Fock state at modulation strengths $K=0$ (solid lines), $K=0.9$ (dashed lines), $K=1.5$ (dot-dashed lines), and $K=3.9$ (dotted lines). (c,d) Derivatives $- d n_{1}^{\rm{c}} / d(Jt)$ of the corresponding curves in (a) and (b), respectively. (e,f) Maximum of the derivative $- d n_{1}^{\rm{c}} / d(Jt)$ in the range $0 \leq Jt \leq (Jt)_{\rm{max}}$ as a function of $K$. For a direct comparison the data sets with $(Jt)_{\rm{max}}<0.5$ are normalized to the value at $K=0$ of the data set with $(Jt)_{\rm{max}}=0.5$. The solid (dashed) line depicts $|J_{0}(K)|$ $(|J_{0}(K)|^2)$ multiplied by the numerically computed maximum of the derivative at $K=0$.}
\end{figure}

In the experiment as discussed in the main text the decay of single occupancy in the presence of modulated interactions is measured in a 3D lattice system. We compare the experimental results presented in Fig.~3(b) of the main article to numerical data extracted from the analysis of the dynamics at $U=0$ within a simplified 3D Bose-Hubbard model including occupation-dependent hopping. Specifically, we investigate the decay of single occupancy on one central site $n_{1}^{\rm{c}}$ that is connected to $6$ neighboring lattice sites in a star-like manner. The corresponding Hamiltonian reads
\begin{equation}
\hat{H}^{\star}_{\rm{eff}} = -J \sum\limits_{\langle i,j \rangle } \hat{a}_i^\dagger J_{0}\left( K (\hat{n}_i - \hat{n}_j) \right) \hat{a}_j + \sum\limits_{i} \frac{U}{2} \hat{n}_i\left(\hat{n}_{i}-1\right) \, ,
\label{SuppEQ1}
\end{equation}
where $\langle i , j \rangle$ denotes nearest neighbors in the star-lattice configuration for a system of $7$ sites in total. Starting from the initial Fock state with one atom on each lattice site, we compute the time evolution driven by $\hat{H}^{\star}_{\rm{eff}}$ for $U=0$ and extract $n_{1}^{\rm{c}}$.

In order to discuss our analysis of the numerical results, let us first consider a simpler scenario where tunneling does not explicitly depend on the occupation numbers. For this, we set the occupation dependence in the argument of the Bessel function $(\hat{n}_i - \hat{n}_j) \equiv 1$, resulting in a simple rescaling of the tunneling rate with $J_0(K)$. The time evolution for $n_{1}^{\rm{c}}$ at different values of $K$ is plotted in Fig.~\ref{FIG8}(a). To identify a consistent measure for the decay rate $\gamma$ from the star-model system, we compute the derivative $- d n_{1}^{\rm{c}} / d(Jt)$, depicted in Fig.~\ref{FIG8}(c), and extract its maximal value within a range $0 \leq Jt \leq (Jt)_{\rm{max}}$. This maximum is plotted in Fig.~\ref{FIG8}(e) as a function of $K$ for different values $(Jt)_{\rm{max}}$. For a direct comparison the results are normalized to the value at $K=0$ of the data set with $(Jt)_{\rm{max}}=0.5$. For very short times, $(Jt)_{\rm{max}} \ll 1$, the maximal derivative scales $\propto |J_0(K)|^2$. However, with increasing $(Jt)_{\rm{max}}$, the maximal derivative follows the expected scaling with the Bessel function $\propto |J_0(K)|$. Therefore, it serves as a good measure for the tunneling rate in the system and thus as a good measure for $\gamma$.

We now consider the full form of $\hat{H}^{\star}_{\rm{eff}}$ including the occupation dependence in the tunneling term. Fig.~\ref{FIG8}(b) shows the corresponding time evolution for $n_{1}^{\rm{c}}$ at different values of $K$. In comparison with the numerical data shown in Fig.~\ref{FIG8}(a), we notice that the occupation-dependent tunneling renders the dynamics more complicated. In particular, the decay in $n_{1}^{\rm{c}}$ is slower for the same $K$. We proceed with the analysis as before and plot $- d n_{1}^{\rm{c}} / d(Jt)$ in Fig.~\ref{FIG8}(d). The extracted maximal derivatives as a function of $K$ for different $(Jt)_{\rm{max}}$ are depicted in Fig.~\ref{FIG8}(f). Again, we find a scaling $\propto |J_0(K)|^2$ at the shortest times. However, with increasing $(Jt)_{\rm{max}}$ the numerical data now lie significantly below a simple scaling with the Bessel function $\propto |J_0(K)|$ owing to the more complex dynamics driven by occupation-dependent tunneling amplitudes. Note that the data sets for $(Jt)_{\rm{max}}=0.15$ and $0.25$ essentially coincide with the data set for $(Jt)_{\rm{max}}=0.5$ and are not shown for clarity. In order to compare the numerical results with the experiment, we rescale the data set for $(Jt)_{\rm{max}}=0.5$ to the measured value for $\gamma$ at $K=0$ and plot it with the experimental data in Fig.~3(b) of the main article. We note that our analysis does not capture possible additional corrections from other sources in the extended 3D lattice, e.g. the presence of a weak harmonic confinement, which are difficult to quantify numerically for larger system sizes.

\begin{figure}
\includegraphics[width=1\columnwidth]{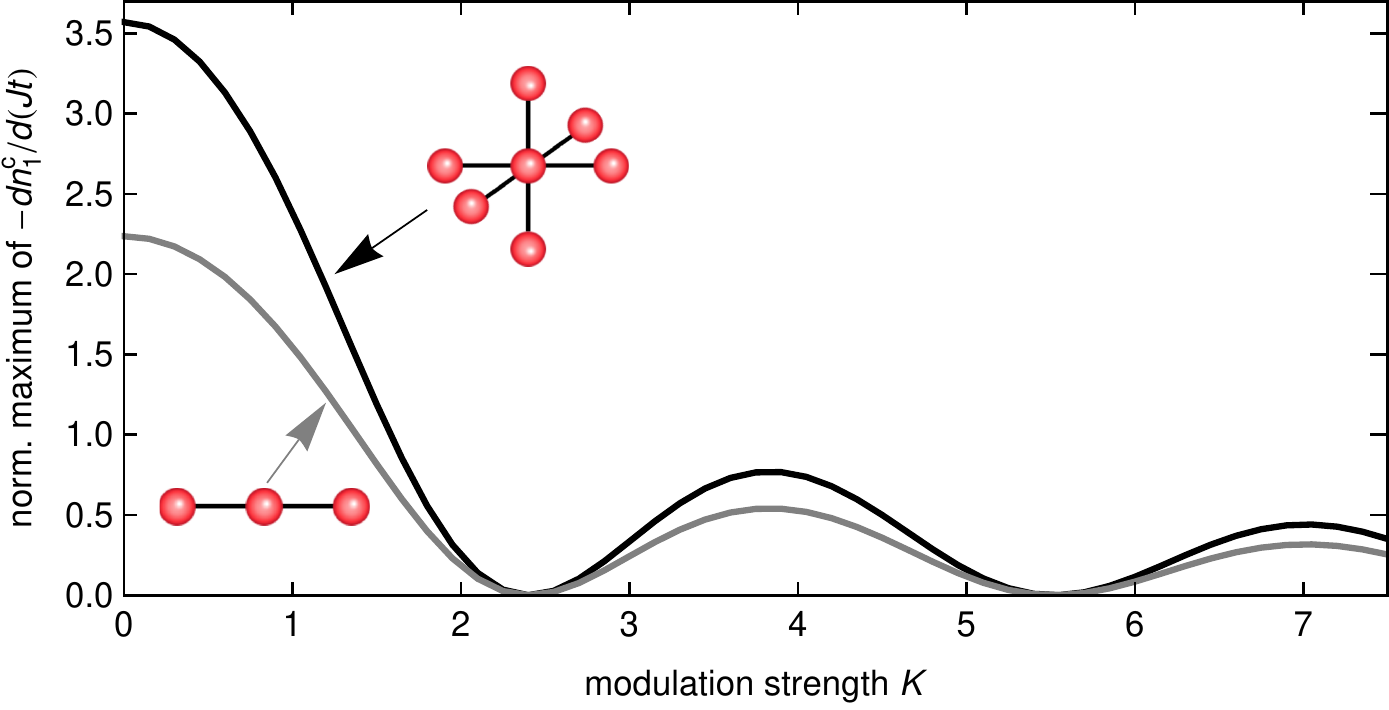}
\caption{\label{FIG9} Comparison of the decay of $n_{1}^{\rm{c}}$ in 1D and 3D. The maximum of the derivative $- d n_{1}^{\rm{c}} / d(Jt)$ with $(Jt)_{\rm{max}}=0.5$ as a function of the modulation strength $K$ is shown for the 3D star-lattice configuration with 7 sites (black line) and the 1D configuration with 3 sites (gray line).}
\end{figure}

Finally, the analysis is repeated for a 1D setting where the central lattice site is connected to only two neighbors. The evaluated maximal derivative for $(Jt)_{\rm{max}}=0.5$ is plotted in Fig.~\ref{FIG9} as a function of $K$ and compared to the 3D case (Fig.~\ref{FIG8}(f)). Generally, we find a faster decay of the single occupancy in the central site $n_{1}^{\rm{c}}$ with increasing dimensionality. This is in agreement with the experimental observation in the large lattice systems, see Fig.~2(c) of the main article. Comparing the values for the maximal derivative in the 3D and 1D configuration at zero modulation strength $K=0$, we obtain a ratio of $1.6$. Within our simplified model system, this ratio provides a quantitative prediction for the measured value $\beta_{\rm{3D}}/\beta_{\rm{1D}}$ reported in the main article.

\bibliographystyle{apsrev}

\clearpage

\end{document}